\documentclass[preprint,nofootinbib]{revtex4}%
\usepackage{amssymb}
\usepackage{amsfonts}
\usepackage{amsmath}
\usepackage{amsmath}
\usepackage{graphicx}
\usepackage[usenames]{color}%
\providecommand{\U}[1]{\protect\rule{.1in}{.1in}}
\providecommand{\U}[1]{\protect\rule{.1in}{.1in}}
\definecolor{blue}{rgb}{0,0,1}

\definecolor{red}{rgb}{1,0,0}

\begin{document}
\title{Exact solutions in $\mathcal{R}^{2}$ SUGRA}
\author{Adolfo Cisterna$^{1}$, Mokhtar Hassa\"{\i}ne$^{2}$ and Julio Oliva$^{3}$.}
\affiliation{$^{1}$Instituto de Ciencias F\'{\i}sicas y Matem\'{a}ticas, Universidad
Austral de Chile, Casilla 567, Valdivia, Chile}
\affiliation{$^{2}$Instituto de Matem\'{a}tica y F\'{\i}sica, Universidad de Talca, Casilla
747, Talca, Chile.}
\affiliation{$^{3}$Departamento de F\'{\i}sica, Universidad de Concepci\'{o}n, Casilla 160-C, Concepci\'{o}n, Chile}

\begin{abstract}
This letter is devoted to show that the bosonic sector of the $\mathcal{R}%
^{2}$-SUGRA in four dimensions, constructed with the F-term, admits a variety
of exact and analytic solutions which include, pp- and AdS waves,
asymptotically flat and AdS black holes and wormholes, as well as product
spacetimes. The existence of static black holes and wormholes implies that a
combination involving the Ricci scalar plus the norm of the field strength of
the auxiliary two-form $B_{\mu\nu}$, must be a constant. We focus on this
sector of the theory which has two subsectors depending on whether such a
combination vanishes or not. \footnote{adolfo.cisterna@uach.cl,
hassaine@inst-mat.utalca.cl, julio.oliva@uach.cl}

\end{abstract}
\maketitle


\section{Introduction}


Recently, there has been renewed interest in theories of gravity with higher
derivative terms, particularly, in the case of theories containing quadratic
terms in the curvature (see e.g. \cite{Lu:2015cqa}-\cite{Lu:2015psa} ).
\newline It is well known that in string theory an infinite series of
higher curvature correction to General Relativity arise naturally. The inclusion of some of these terms makes the gravitational
theory under consideration renormalizable but the truncation of the series to
an effective field theory is done at the cost of introducing a massive ghost
degree of freedom. At quadratic order, this state appears due to the presence
of terms which are the square of the Riemann, Ricci or Weyl
tensors\footnote{At the level of the field equations in $D=4$, only two of
these are independent since the Euler density does not contribute to the
equations of motion.} \cite{stelle renorm}. \newline The case of $R+R^{n}$
gravity is particularly interesting, since after a field redefinition it can
be rewritten as General Relativity with a minimally coupled scalar field with
a self interaction. In the context of cosmology Starobinsky considered the
simplest modification to GR, the case of $n=2$, and it was shown that the
effective potential for the scalar degree of freedom has a structure that is
very ad hoc for the study of cosmic inflation. Indeed, it describes a slow roll
transition from a de Sitter phase to a flat Minkowski phase \cite{starmodel},
leading to a possible realization of an inflationary era in the early
Universe. Even more, cosmological scenarios have been investigated embedding
this model in the context of supergravity theories (see e.g.
\cite{Kounnas:2014gda}). \newline Moreover, contrary to other higher
derivative extensions, $R+R^{2}$ gravity is ghost-free. The theory along with
a massless graviton also propagates a massive scalar degree of freedom usually
called scalaron field \cite{Alvarez-Gaume:2015rwa}. Even though in higher
derivative modified gravity the Einstein term is usually present, recently the
case of pure higher derivative theories have been investigated in various
contexts \cite{Kehagias:2015ata}, \cite{Deser:2007vs} and different features of black hole thermodynamics have been explored in \cite{Cognola:2015wqa}, \cite{Cognola:2015uva}. The specific case of
$R^{2}$ theory shows interesting properties. $R^{2}$ gravity is ghost free and
scale invariant. The scale invariance holds classically and it is expected
that at quantum level the Einstein-Hilbert term is induced by broking the
symmetry in a soft manner \cite{Kounnas:2014gda}. The theory is conformally
equivalent to general relativity plus cosmological constant with minimally
coupled scalar fields and its particle content depends on the background. For
instance on a flat background the theory only propagates a scalar mode while
in a curved background (de Sitter), along with the scalar state it propagates
a massless graviton. The later is in agreement with the fact that for $R\neq0$
the theory is equivalent as we mentioned above to general relativity with
cosmological constant plus minimally coupled scalar fields. For $R=0$ this
conformal transformation becomes singular.\newline Many of these features let
the authors of \cite{Ferrara:2015ela} to revisit the supersymmetric extensions
of pure $R^{2}$ gravity. In its new minimal formulation, the multiplet of the
$R^{2}$ supergravity consists on a vierbein and its supersymmetric partner the
gravitino, and the off-shell realization of supersymmetry is ensured by
introducing two auxiliary fields that turn out to be a gauge field together
with a two-form, both possessing the gauge symmetry. \newline The aim of this
letter is to explore the space of solutions of the bosonic sector of this
$\mathcal{R}^{2}$ supergravity. We will first establish that the existence of
regular static solutions imposes that a particular linear combination of the
scalar curvature together with the norm of the dual field strength associated
to the two-form must be a constant $c$. Under this restriction, the original
field equations reduce to a second-order, traceless system. For a constant
$c=0$, the theory is no longer equivalent to a scalar tensor theory and the
Einstein field equations are automatically satisfied by imposing a certain
relation between the gauge field and the dual field strength. In this case, we
will construct black holes and wormholes. At this particular point, pp-waves
are also admissible with a null dual field. On the other hand for $c\not =0$,
we establish the existence of product spaces of the form $\mathbb{R}\times
H_{3}$ and $dS_{3}\times\mathbb{R}$ where $H_{3}$ and $dS_{3}$ denote
respectively the three-dimensional hyperbolic and de-Sitter spaces.
Interestingly enough, the field equations are also shown to support AdS-waves
or Siklos spacetimes.

This paper is organized as follows: In Section II, we define the theory,
providing the Lagrangian as well the field equations. In Section III, we
establish that the existence of static regular solutions requires that the
following combination $R+6H_{\mu}H^{\mu}$ must be a constant $c$, where $R$
stands for the scalar curvature and $H_{\mu}$ denotes the dual field strength.
This latter condition notably simplifies the system for asymptotically flat or
AdS black holes and wormholes, reducing the field equations to a traceless,
second order system. Section IV is devoted to present some interesting
solutions with non-vanishing $c$ while in Section V we construct static black
holes and wormholes that correspond to solutions with $c=0$. In the last
Section, we provide some further comments.


\section{The theory}

In this section, we present the bosonic sector of the $F-$terms of the
$\mathcal{R}^{2}$ supergravity (see \cite{Ferrara:2015ela}). The bosonic
action is described by the metric and the two auxiliary gauge fields, and is
given by\footnote{As shown in \cite{Ferrara:2015ela}, generically, this theory is conformally equivalent to a second order theory, and therefore free of Ostrogradski instabilities.}
\begin{equation}
I\left[  g_{\mu\nu},A_{\mu},B_{\mu\nu}\right]  =\int\sqrt{-g}d^{4}x\left(
\frac{1}{8g^{2}}\left(  R+6H_{\mu}H^{\mu}\right)  ^{2}-\frac{1}{4g^{2}}%
F_{\mu\nu}\left(  A_{\rho}^{-}\right)  F^{\mu\nu}\left(  A_{\rho}^{-}\right)
\right)  \ , \label{action}%
\end{equation}
where
\begin{equation}
A_{\mu}^{-}=A_{\mu}-3H_{\mu}\ ,
\end{equation}
and
\begin{equation}
H_{\mu}=-\frac{1}{3!}\epsilon_{\mu\nu\rho\sigma}H^{\nu\rho\sigma}.
\end{equation}
Here, the three-form $H_{\mu\nu\lambda}$ is defined as the field strength of
the auxiliary two form $B_{\mu\nu}$, i.e., $H_{\nu\rho\sigma}:=\partial_{\mu
}B_{\nu\rho}$ + cyclic perm. The local existence of $B_{\mu\nu}$ is ensured
provided the condition $\nabla_{\mu}H^{\mu}=0$ holds.

Note that the action is invariant under the rigid scale transformation
\begin{equation}
g_{\mu\nu}\rightarrow\omega g_{\mu\nu}\text{, }\quad A_{\rho}\rightarrow
A_{\rho}\text{, }\quad H_{\mu}\rightarrow H_{\mu}, \label{scaleinvariance}%
\end{equation}
where $\omega$ is an arbitrary non-vanishing constant.

The corresponding field equations coming from the variations of the action
(\ref{action}) with respect to $g_{\mu\nu}$, $B_{\mu\nu}$ and $A_{\mu}$,
reduce to the following set of equations%

\begin{equation}
\left[  g_{\mu\nu}\square-\nabla_{\mu}\nabla_{\nu}+R_{\mu\nu}-\frac{1}%
{4}g_{\mu\nu}R+6\left(  H_{\mu}H_{\nu}-\frac{1}{4}g_{\mu\nu}H_{\alpha
}H^{\alpha}\right)  \right]  \left(  R+6H_{\alpha}H^{\alpha}\right)
=2T_{\mu\nu}^{M}\left(  A^{-}\right)  \ , \label{eg}%
\end{equation}%
\begin{align}
\nabla^{\mu}[(R+6H_{\alpha}H^{\alpha})H_{\mu\beta\gamma}]  &  =0\ ,
\label{eh}\\
\nabla_{\mu}F^{\mu\nu}\left(  A^{-}\right)   &  =0\ , \label{ea}%
\end{align}
with $T_{\mu\nu}^{M}$ being the Maxwell energy-momentum tensor defined as
\begin{equation}
T_{\mu\nu}^{M}\left(  A^{-}\right)  :=F_{\mu\alpha}\left(  A^{-}\right)
F_{\nu}^{\ \ \alpha}\left(  A^{-}\right)  -\frac{1}{4}g_{\mu\nu}F_{\alpha
\beta}\left(  A^{-}\right)  F^{\alpha\beta}\left(  A^{-}\right)  \ .
\end{equation}

In the next section we will establish that the existence of regular, static
solutions requires the following combination $R+6H_{\alpha}H^{\alpha}$ to be a constant.


\section{On static and regular solutions}

Let us consider the following static ansatz for the metric
\begin{equation}
ds^{2}=-\lambda^{2}\left(  x\right)  dt^{2}+h_{ij}\left(  x\right)
dx^{i}dx^{j}\ , \label{1mas3}%
\end{equation}
where $h_{ij}\left(  x\right)  $ is a spacelike metric and $\lambda\left(
x\right)  $ depends only on the spacelike coordinates $x^{i}$. This metric may
cover the domain of outer communications of a static black hole or even a
global metric on a wormhole geometry. In the former case, the function
$\lambda\left(  x\right)  $ vanishes at the horizon while for the wormhole
solution the function is non-vanishing everywhere. Considering the trace of
equation (\ref{eg}) and multiplying the result by $\lambda\left(  x\right)
\left(  R+6H_{\mu}H^{\mu}\right)  $ one obtains%
\begin{equation}
D_{i}\left(  \lambda\left(  x\right)  \left(  R+6H_{\mu}H^{\mu}\right)
D^{i}\left(  R+6H_{\mu}H^{\mu}\right)  \right)  -\lambda\left(  x\right)
D_{i}\left(  R+6H_{\mu}H^{\mu}\right)  D^{i}\left(  R+6H_{\mu}H^{\mu}\right)
=0\ ,
\end{equation}
where $D_{i}$ is the covariant derivative intrinsically defined by
$h_{ij}\left(  x\right)  $. Integrating this result on a $t$-constant
(spacelike) hypersurface and disregarding boundary terms one obtains%
\begin{equation}
R+6H_{\mu}H^{\mu}=c\ , \label{constraintc}%
\end{equation}
where $c$ is a constant. To be able to drop the boundary terms, in the black
hole geometry we have to require a certain asymptotic behavior at infinity,
while to do so in the case of the wormhole geometry we have to require such
asymptotic behavior to be valid at both asymptotic regions.

Note that by performing a rigid scale transformation, according to the
symmetry (\ref{scaleinvariance}), we can normalize the constant $c$ to $\pm1$
or $0$.

The constraint (\ref{constraintc}) considerably simplifies the original field
equations (\ref{eg}) and it is clear that the cases $c=0$ and $c\not =0$ must
be treated separately. Indeed, for a non-vanishing $c$, the equations
(\ref{eg}-\ref{ea}) yields a reduced system defined by the following second
order equations
\begin{align}
&  R_{\mu\nu}-\frac{1}{4}g_{\mu\nu}R+6\left(  H_{\mu}H_{\nu}-\frac{1}
{4}g_{\mu\nu}H_{\alpha}H^{\alpha}\right)  =\frac{2}{c}T_{\mu\nu}^{M}\left(
A^{-}\right)  \ ,\label{egred}\\
&  R+6H_{\mu}H^{\mu} =c\ ,\label{const}\\
&  \nabla_{\mu}H^{\mu\beta\gamma}=0\ ,\nabla_{\mu}F^{\mu\nu}\left(
A^{-}\right)  =0\ , \label{max}%
\end{align}
with the further requirement that $\nabla_{\mu}H^{\mu}=0$.

Even though this reduced system is of second order, it is non-trivial to find
general solutions for spacetimes with a given symmetry. Nevertheless we have
been able to construct some scattered solutions that we report and analyze in
the next sections. On the other hand, if the constant $c$ vanishes, besides
the gravitational constraint $R+6H_{\mu}H^{\mu}=0$, one has that the theory
reduces to Maxwell theory for $A_{\mu}^{-}$ with the further constraint of
vanishing energy-momentum tensor, i.e.%
\begin{equation}
\nabla_{\mu}F^{\mu\nu}\left(  A^{-}\right)  =0\ ,\text{ }T_{\mu\nu}^{M}\left(
A^{-}\right)  =0\ .
\end{equation}
For the static ansatz (\ref{1mas3}), this implies that $A_{\mu}$ and $3H_{\mu
}$ are equal, up to a total derivative.

\bigskip

We separate the solutions we have found in two families depending on whether
$c$ in (\ref{constraintc}) vanishes or not.


\section{$AdS$ waves and product spaces with non-vanishing $c$}


We now turn our analysis in looking for solutions of the reduced field
equations (\ref{egred})-(\ref{const})-(\ref{max}) for which the constant
$c\not =0$. Since AdS-waves as well as products of constant curvature
spacetimes have constant scalar invariants, it is natural to explore the
possibility for such spacetimes to be solutions of our reduced field
equations. In addition, we assume that the $U(1)$ gauge field $A_{\mu}$ and
the dual field $H_{\mu}$ are proportional
\begin{equation}
A_{\mu}=3H_{\mu},
\end{equation}
condition which automatically ensures that the Maxwell equations are
satisfied, and also allows to decouple the Maxwell part from the Einstein
equations. Indeed, in this case the latter reduce
\begin{equation}
R_{\mu\nu}-\frac{1}{4}g_{\mu\nu}R+6\left(  H_{\mu}H_{\nu}-\frac{1}{4}g_{\mu
\nu}H_{\alpha}H^{\alpha}\right)  =0\ , \label{redeqs}%
\end{equation}
and where the divergence-free condition $\nabla_{\mu}H^{\mu}=0$ must also hold.

$\bullet$ The product spaces $\mathbb{R}\times H_{3}$:\newline Let us consider
the product spaces $\mathbb{R}\times H_{3}$ where $H_{3}$ denotes the
three-dimensional hyperbolic space. In this case, a solution is given by
\begin{equation}
ds^{2}=-dt^{2}+\frac{l^{2}}{y^{2}}\left(  dx^{2}+dy^{2}+dz^{2}\right)  ,\qquad
H=\pm\frac{\sqrt{3}}{3l}dt.
\end{equation}
This corresponds to a solution of (\ref{redeqs}) with $c=-8/l^{2}$, and since
the equations are local, one could as well consider quotients $H_{3}/\Gamma$,
where $\Gamma$ is a freely acting, discrete subgroup of $SO\left(  3,1\right)
$. This identification might have some effect on the possible global existence
of Killing spinors.

$\bullet$ The product spaces $dS_{3}\times\mathbb{R}$:\newline As before, we
show that the product of the three-dimensional de Sitter space with
$\mathbb{R}$ provide a simple solution of the reduced equations with
non-vanishing $c$. Indeed, in this case, the solution reads
\begin{equation}
ds^{2}=-\left(  1-\frac{r^{2}}{l^{2}}\right)  dt^{2}+\frac{dr^{2}}%
{1-\frac{r^{2}}{l^{2}}}+r^{2}d\phi^{2}+dz^{2},\qquad H=\pm\frac{\sqrt{3}}%
{3l}dz,
\end{equation}
and corresponds to a solution with a positive constant $c=8/l^{2}$. Note that
this solution can be Wick rotated yielding a new Euclidean solution which
reduces locally to $S^{3}\times\mathbb{R}$. The fields $H$ and $A$ remain
invariant under the Wick rotation.

$\bullet\ $AdS-waves:\newline As mentioned before, the AdS waves present the
advantage that all their scalar invariants are constants making these
confirgurations possible interesting candidates for the sector of the theory
we are exploring. Note that such configurations have been shown to be
solutions of Einstein gravity with a nonminimal scalar field
\cite{AyonBeato:2005qq} as well as for quadratic corrections of the Einstein
gravity \cite{AyonBeato:2009yq,AyonBeato:2011qw}. The line element of the AdS
waves can be parameterized as follows
\begin{equation}
ds^{2}=\frac{l^{2}}{y^{2}}\left(  -F(u,y,x)du^{2}+2dudv+dy^{2}+dx^{2}\right)
, \label{adswaveans}%
\end{equation}
which clearly emphasize that these spacetimes can be obtained from AdS through
a Kerr-Schild transformation. This allows to interpret the AdS waves as exact
gravitational waves propagating on the AdS space, and consequently their
matter source must behave as a pure radiation field. This last condition may
be satisfied requiring the dual field $H_{\mu}$ to be along the retarded time
$u$ as
\begin{equation}
H_{\mu}dx^{\mu}=f(u,y,x)du,
\label{dfu}
\end{equation}
where $f$ is a function depending on all the coordinates except the null one
$v$. Note that the divergenceless condition of $H_{\mu}$ is automatically
satisfied in this case. The field equations (\ref{redeqs}) with an Ansatz of
the form (\ref{adswaveans}-\ref{dfu}) will be fulfilled provided that the
following equation
\begin{equation}
\frac{\partial^{2}F}{\partial x^{2}}+\frac{\partial^{2}F}{\partial y^{2}%
}-\frac{2}{y}\frac{\partial F}{\partial y}+12f^{2}=0\ ,
\label{eqF}
\end{equation}
holds. In this case $R=-12/l^{2}$ and since $H_{\mu}$ is null the constant
$c=-12/l^{2}$ is non-vanishing. This means that for any null field of the form
(\ref{dfu}), one is able in principle to find AdS wave configurations provided
that the structural function $F$ satisfies the equation (\ref{eqF}). For
example, in the case with $f$ being a constant, this equation may be easily
integrated yielding
\begin{equation}
F=a\left(  u\right)  \left(  x^{2}+y^{2}\right)  +b\left(  u\right)
x+c\left(  u\right)  y^{3}+6f^{2}y^{2}+d\left(  u\right)  \ ,
\end{equation}
where $a$, $b$, $c$ and $d$ are arbitrary functions of the retarded time $u$.

Just to conclude this section, we would like to mention that pp waves are also
solutions of the equations (\ref{eqF}) with a null dual field, but in this
case they correspond to a solution with a vanishing constant $c$.

\bigskip

The solutions we have just mentioned have non-vanishing $c$, and we have
presented cases with $c$ being positive an negative. In the next section we
present solutions with vanishing $c$.


\section{Solutions with vanishing $c$. Black holes, wormholes and beyond}

We now look for solutions for which the integration constant $c=0$, and we
first consider the following spherically symmetric Ansatz
\begin{equation}
ds^{2}=-f\left(  r\right)  dt^{2}+\frac{dr^{2}}{f\left(  r\right)  }+\left(
r^{2}+a^{2}\right)  d\Omega_{2}^{2}\ , \label{metricblackworm}%
\end{equation}
where $d\Omega_{2}$ stands for the line element of the two-sphere. If
$f\left(  r\right)  $ is non-vanishing, this metric describes a wormhole
geometry where the surface $r=0$ represents a traversable wormhole throat.
Geodesics with vanishing angular momentum are pulled towards the minimum of
the function $f\left(  r\right)  $. On the other hand, if $f\left(  r\right)
$ vanishes the metric (\ref{metricblackworm}) may represent a black hole
geometry with an event horizon located at $r=r_{+}$ such that $f\left(
r_{+}\right)  =0$.

The whole system of equations (\ref{eg}-\ref{ea}) are solved by
\begin{equation}
A_{\mu}=3H_{\mu}\text{, }H=\frac{\chi(\theta)}{r^{2}+a^{2}}dt+\frac
{\chi(\theta)}{f(r)(r^{2}+a^{2})}dr+\frac{\sqrt{2}\sqrt{r^{2}+a^{2}}}{l}%
\sin(\theta)d\phi\ ,
\end{equation}
provided%
\begin{equation}
\frac{d^{2}f\left(  r\right)  }{dr^{2}}+\frac{4r}{r^{2}+a^{2}}\frac{df\left(
r\right)  }{dr}+\frac{2\left(  r^{2}+2a^{2}\right)  }{\left(  r^{2}%
+a^{2}\right)  ^{2}}f\left(  r\right)  =\frac{2\left(  6r^{2}+6a^{2}%
+l^{2}\right)  }{l^{2}\left(  r^{2}+a^{2}\right)  }\ . \label{mef}%
\end{equation}
Here $H_{\mu}H^{\mu}=\frac{2}{l^{2}}$ and $H_{\mu}$ is divergenceless,
ensuring the local existence of the auxiliary two-form $B_{\mu\nu}$. The
function $\chi(\theta)$ is arbitrary, and in this case the constant $c$ in
equation (\ref{constraintc}) vanishes.

The asymptotic expansion of (\ref{mef}) at $r\rightarrow\pm\infty$, reveals
that the leading term in $f\left(  r\right)  $ behaves as $f\left(  r\right)
\rightarrow\frac{r^{2}}{l^{2}}$, and therefore the solutions of (\ref{mef})
will provide us with asymptotically locally AdS spacetimes. The general
solution of (\ref{mef}) can be written in terms of hypergeometric functions
$_{2}F_{1}$, and is given by{\small
\begin{align*}
f(r)  &  =\frac{e^{-i\sqrt{3}\tan^{-1}\left(  \frac{r}{a}\right)  }}%
{6a(a^{2}+r^{2})^{3/2}}\left(  -(a^{2}+r^{2})\left(  6ae^{i\sqrt{3}\tan
^{-1}\left(  \frac{r}{a}\right)  }\sqrt{a^{2}+r^{2}}\left(  4a^{2}%
+r^{2}+1+c_{1}\frac{e^{i\sqrt{3}\tan^{-1}\left(  \frac{r}{a}\right)  }}%
{\sqrt{a^{2}+r^{2}}}\right)  +\sqrt{3}ic_{2}\right)  \right. \\
&  +2a^{2}(1+3a^{2})e^{i\sqrt{3}\tan^{-1}\left(  \frac{r}{a}\right)
}(a+ir)\sqrt{a^{2}+r^{2}}\left(  (3+\sqrt{3})\ _{2}F_{1}\left[  1,\frac{1}%
{2}(1-\sqrt{3}),\frac{1}{2}(3-\sqrt{3}),1-\frac{2a}{a-ir}\right]  \right. \\
&  \left.  \left.  -(-3+\sqrt{3})\ _{2}F_{1}\left[  1,\frac{1}{2}(1-\sqrt
{3}),\frac{1}{2}(3-\sqrt{3}),1-\frac{2a}{a-ir}\right]  \right)  \right)  \ ,
\end{align*}
} where $c_{1}$ and $c_{2}$ are two integration constants. Since the equation
(\ref{mef}) is a linear, non-homogeneous, ordinary differential equation, one
can take the real part of this solution to obtain a function that defines a
metric. Depending on the values of the integration constants the solution may
describe black hole or wormhole configurations as shown in Figure 1. These
solutions extend the solutions recently found in \cite{Duplessis:2015xva} to
the asymptotically AdS case.

\begin{figure}[h]
\includegraphics[scale=0.5]{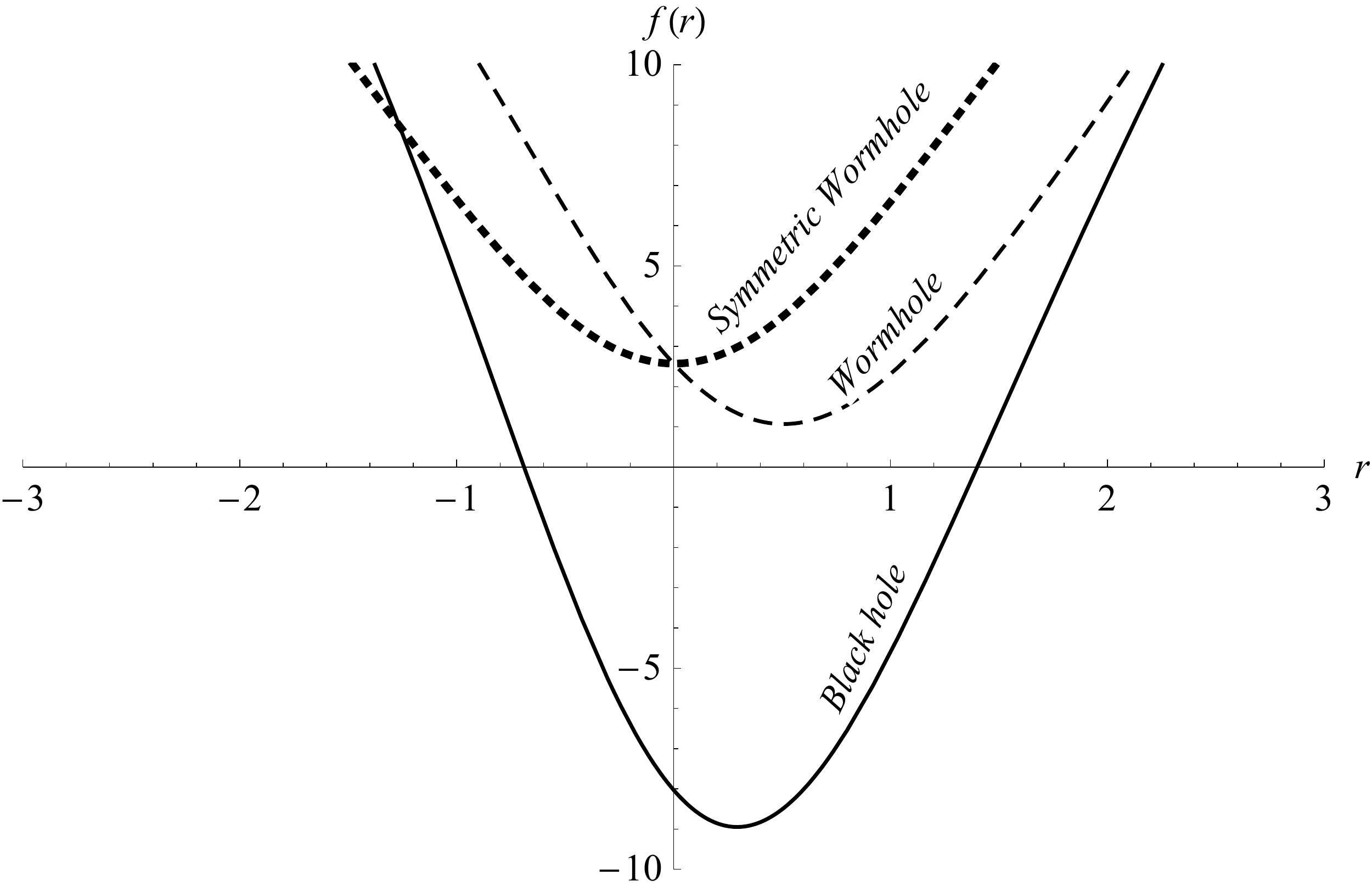}
\caption{Real part of $f(r)$ vs $r$ for different values of the integration constants.}
\label{Tvsrmas}
\end{figure}

The black line in Figure 1, describes a regular black hole, which is
asymptotically AdS. The first zero of the function $f\left(  r\right)  $ (from
right to left) defines the event horizon, while the second one defines a
Cauchy horizon. The dashed line in Figure 1, describes a wormhole which is
asymptotically AdS at both asymptotic\ regions. The lapse function has a
minimum at a certain point which in general differs from the location of the
throat (located at $r=0$). As observed for example in \cite{Dotti:2007az} and
\cite{Duplessis:2015xva}, this has an interesting effect on the geodesics with
angular momentum since gravity pulls geodesics with vanishing angular momentum
towards the minimum of the function $f\left(  r\right)  $ while the
centrifugal contribution reverses its direction at the throat. Therefore,
there is a region between the throat and the minimum of the function $f\left(
r\right)  $ on which both contributions on geodesics with angular momentum
point in the same direction and therefore the geodesic flow obligates a
particle to go out from such region.

\bigskip

As a second example of solutions with $c=0$, one can consider as well the most
general Petrov type D spacetime in four dimensions which is the
Plebanski-Demianski spacetime, whose metric is given by%
\begin{equation}
ds^{2}=\frac{1}{\left(  1-pq\right)  ^{2}}\left[  \left(  p^{2}+q^{2}\right)
\left(  \frac{dp^{2}}{X\left(  p\right)  }+\frac{dq^{2}}{Y\left(  q\right)
}\right)  +\frac{X\left(  p\right)  }{p^{2}+q^{2}}\left(  d\tau+q^{2}%
d\sigma\right)  ^{2}-\frac{Y\left(  q\right)  }{p^{2}+q^{2}}\left(
d\tau-p^{2}d\sigma\right)  ^{2}\right]  \ .
\end{equation}
In this case, one can show that the system (\ref{eg}-\ref{ea}) is fulfilled
for this metric with%
\begin{equation}
H=\frac{\left(  pq-1\right)  }{l}\sqrt{\frac{2X(p)}{p^{2}+q^{2}}}%
dt+\frac{q^{2}}{l}\sqrt{\frac{2X\left(  p\right)  }{\left(  p^{2}%
+q^{2}\right)  \left(  1-pq\right)  ^{2}}}d\sigma\ ,
\end{equation}
where $X(p)$ is positive definite. Here $H_{\mu}H^{\mu}=2/l^{2}$ and the
metric functions $X$ and $Y$ are given by
\begin{align}
X\left(  p\right)   &  =x_{0}+x_{1}p+x_{2}p^{2}+x_{3}p^{3}+x_{4}p^{4}\ ,\\
Y\left(  q\right)   &  =-\left(  x_{4}-\frac{1}{l^{2}}\right)  -x_{3}%
q-x_{2}q^{2}-x_{1}q^{3}-\left(  x_{0}-\frac{1}{l^{2}}\right)  q^{4}\ .
\end{align}


\section{Further comments}

Here, we have considered the bosonic sector of the $\mathcal{R}^{2}$-SUGRA in
four dimensions for which we have obtained a variety of exact and analytical
solutions. We have first shown that requiring the existence of regular static
solutions, the space of configurations is such that the following combination
$R+6H_{\alpha}H^{\alpha}$ must be a constant $c$.

Solutions with non-vanishing $c$ are provided by AdS waves or product spaces
of the form $\mathbb{R}\times H_{3}$ and $dS_{3}\times\mathbb{R}$. At this
point, one may note a certain analogy with the standard eleven-dimensional
supergravity whose purely bosonic field equations, given by
\[
R_{\mu\nu}-\frac{1}{2}g_{\mu\nu}R=\frac{1}{12}\Big(F_{\mu\nu}^{2}-\frac{1}%
{8}g_{\mu\nu}F^{2}\Big),
\]
are quite similar to our reduced field equations (\ref{redeqs}). Here $F$ is
the field strength associated to the three-form with $F_{\mu\nu}^{2}%
=F_{\mu\alpha\beta\sigma}F_{\nu}^{\,\alpha\beta\sigma}$ and $F^{2}%
=F_{\mu\alpha\beta\sigma}F^{\mu\alpha\beta\sigma}$. This analogy with the
standard eleven-dimensional supergravity can also be extended to the spectrum
of solutions where our product space solutions are similar to the Freund-Rubin
solutions $AdS_{4}\times S_{7}$ and $AdS_{7}\times S_{4}$ of
eleven-dimensional supergravity. A natural task to explore would be to
determine the amount of supersymmetry preserved by these backgrounds by
solving the Killing spinorial equation, which is explicitly given in \cite{Ferrara:1988qxa}. We leave this task for future work.   

\bigskip
\textbf{Acknowledgments}
This work has been supported by FONDECYT Regular grants 1141073, 1150246 and
1130423. This project was also partially funded by Proyectos CONICYT, Research
Council UK (RCUK) Grant No. DPI20140053. A.C. work is supported by FONDECYT
project N\textordmasculine3150157.

\end{document}